\documentclass[prapplied,floatfix,reprint,superscriptaddress]{revtex4-2}

\usepackage{xcolor}
\usepackage{amsmath}
\usepackage{graphicx}

\def\n2x{n_{2,\langle 100 \rangle}}
\def\alphaR{\alpha_R}
\def\bE{\mathbf{E}}
\def\bP{\mathbf{P}}

\begin{document}
\title{Raman-induced Kerr effect and heating in diamond}

\author{C. D. Cruz}
\affiliation{Nanoscale Device Characterization Division, National Institute of Standards and Technology, Gaithersburg, MD, 20899}
\author{J. C. Stephenson}
\affiliation{Associate, Sensor Science Division, National Institute of Standards and Technology, Gaithersburg, MD, 20899}
\author{J. K. Wahlstrand}
\email{jared.wahlstrand@nist.gov}
\affiliation{Nanoscale Device Characterization Division, National Institute of Standards and Technology, Gaithersburg, MD, 20899}

\begin{abstract}
We fully characterize the polarization-dependent third-order nonlinear susceptibility in a diamond crystal using phase-sensitive nondegenerate pump-probe spectroscopy near 1~$\mu$m wavelength.
Measuring the frequency-domain phase shift and amplitude change of the probe pulse as a function of polarization and probe wavelength allows direct extraction of the full anisotropic $\chi^{(3)}$ response, including the Raman contribution.
We find that the polarization dependence of the two-beam coupling contribution for the Raman component differs from that predicted solely using one-beam selection rules.
We measure a Kerr coefficient that is within the range of reported values in the literature, and our measured anisotropy parameters satisfy Kleinman symmetry, as expected.
Our measured Raman polarizability agrees, after scaling for its dispersion, with a Brillouin-referenced measurement at 514.5~nm [Grimsditch and Ramdas, Phys. Rev. B \textbf{11}, 3139 (1975)], but is lower than measurements from absolute Raman cross sections.
The same technique could be used to measure other materials with relatively high phonon frequencies, and it complements measurements of the Kerr contribution of low-frequency modes using impulsive stimulated Raman scattering (ISRS).
When the pump-probe difference frequency matches the phonon resonance, Raman-enabled two-photon absorption occurs; we directly measure the resulting heat deposition in diamond using time-resolved beam deflection of a continuous-wave laser.
The measurement technique demonstrated here is capable of precise absolute $\chi^{(3)}$ measurements and will enable quantitative stimulated Raman heating microscopy, with potential extensions to thermal conductivity metrology.
\end{abstract}

\maketitle

\section{Introduction}

The optical Kerr coefficient $n_2$, which characterizes the irradiance-dependent refractive index, arises from both the bound electronic third-order response and non-instantaneous processes, such as stimulated Raman scattering, that are often folded into the effective nonlinear susceptibility $\chi^{(3)}$.
Because the Raman response is time-dependent and obeys distinct selection rules, its effective magnitude varies with optical pulse duration, shape, and polarization.
In solids, this contribution to $\chi^{(3)}$ has been probed using ``three-wave mixing'' \cite{levenson_interference_1972,levenson_dispersion_1974}, absolute measurements of Raman scattering cross sections \cite{hellwarth_origin_1975}, and impulsive stimulated Raman scattering (ISRS) \cite{smolorz_time-resolved_1998,smolorz_measurement_1999}.
Its delayed temporal behavior directly influences soliton formation, making it a critical parameter in the design and modeling of integrated photonic devices \cite{cherenkov_raman-kerr_2017}.
Furthermore, this sensitivity to pulse parameters and polarization likely contributes to the scatter among $n_2$ measurements in materials such as glasses, where the Raman response is relatively strong. This response is also highly sensitive to structural changes in amorphous materials, which further complicates comparison across samples. Achieving measurement uncertainties below 10~\% in these materials will require careful accounting of the Raman response.

Single-beam methods for characterizing the Kerr response can access its time dependence, and therefore the Raman contribution, by measuring the dependence on pulse duration \cite{ganeev_nonlinear_2004}. The polarization dependence can also be used to bound the Raman contribution \cite{schiek_nonlinear_2023}; however, both of these techniques offer limited precision.
Two-beam methods that separate excitation from detection are better suited to precise measurements, and pump-probe techniques have been developed based on spectrally resolved two-beam coupling \cite{smolorz_measurement_1999}, spectral interferometry \cite{wahlstrand_absolute_2012}, and beam deflection \cite{reichert_temporal_2014}.
These methods are all based on ISRS. 
Resolving the Raman response temporally with them requires pulses shorter than the time scale of the underlying Raman dynamics, and the strength of the ISRS response can be sensitive to the pulse shape \cite{yan_impulsive_1987,merlin_generating_1997}.
Because the shape of ultrashort pulses evolves during propagation due to dispersion and, potentially, nonlinear interactions, using sufficiently short pulses complicates the interpretation and numerical modeling of ISRS-based measurements, particularly for high-frequency modes.

Here, we employ a phase-sensitive pump-probe spectroscopy method, first demonstrated in H$_2$ and D$_2$ gases \cite{wahlstrand_absolute_2015}, that provides both spectral and temporal information on the Raman contribution to $n_2$.
This technique is well suited to high-frequency phonon modes that are difficult to measure using ISRS and, to our knowledge, has not previously been applied to solids.
We apply it to diamond, whose Raman response is strong and simple, with a single peak at 39.9~THz (1332~cm$^{-1}$), a very high frequency for a solid.
Phase-sensitive pump-probe spectroscopy enables direct measurement of both the Raman and the bound electronic contributions, allowing full characterization of the anisotropic bound electronic $\chi^{(3)}$ tensor and the Raman polarizability, which determines the strength of Raman scattering.
The imaginary part of the effective nonlinear susceptibility gives the Raman contribution to the two-photon absorption coefficient.
This absorption deposits heat directly into the lattice and is connected to recently developed Raman heating techniques for microscopy \cite{zhu_stimulated_2023,zhu_stimulated_2026}.
It also affects the performance of high-power Raman lasers \cite{mildren_highly_2009,antipov_12_2019}.
Finally, we use our measurements to calculate the Raman-mediated nonlinear absorption and compare the result with a beam-deflection measurement of the thermal response in the same diamond sample.

\section{Theory of the third-order response in nondegenerate pump-probe experiments}

We treat the electronic and Raman contributions separately and sum them to find the total response.

\subsection{Bound electronic contribution to $n_2$}
The theory of the electronic $\chi^{(3)}$ polarization dependence is well developed for zincblende and diamond crystals \cite{dvorak_measurement_1994,hutchings_theory_1995}.
In these cubic crystals, there are three independent tensor components for nearly degenerate wavelengths, $\chi_{xxxx}$, $\chi_{xyxy}$, and $\chi_{xxyy}$, which are often expressed in terms of three alternative parameters.
For the real part of the susceptibility (relevant here because we use near-infrared light in diamond, for which electronic two-photon absorption is forbidden energetically), these are the Kerr coefficient for a linearly polarized field along $\langle 001 \rangle$, $\n2x = 3\,\mathrm{Re}\,\chi_{xxxx}/(4n_0^2c\varepsilon_0)$, where $c$ is the speed of light, $\varepsilon_0$ is the vacuum permittivity, and $n_0$ is the linear refractive index; the anisotropy parameter, $\sigma = \mathrm{Re}\,[(\chi_{xxxx} - \chi_{xxyy} - 2\chi_{xyxy})/\chi_{xxxx}]$; and the incremental circular dichroism parameter, $\delta = \mathrm{Re}\,[(\chi_{xxxx} + \chi_{xxyy} - 2\chi_{xyxy})/(2\chi_{xxxx})]$.
The response for any light polarization with respect to the crystal axes can be expressed in terms of these three parameters.

In terms of $\n2x$, $\sigma$, and $\delta$, the electronic Kerr coefficients for linearly polarized pump and probe beams with parallel ($n_{2,\parallel,e}$) and perpendicular ($n_{2,\perp,e}$) polarizations are, for our (001)-oriented crystal with the pump polarization oriented at an angle $\theta$ from the [100] direction,
\begin{align}
n_{2,\parallel,e} (\theta) &= 2\n2x \left(1-\frac{\sigma}{2} \sin^2 2\theta \right), \label{elec_par} \\
n_{2,\perp,e} (\theta) &= \n2x \left(1-\delta-\frac{\sigma}{2}+\sigma \sin^2 2\theta\right), \label{elec_perp}
\end{align}
where we have included the factor of 2 (cross-phase modulation) enhancement of the nonlinear response in a pump-probe experiment \cite{wahlstrand_effect_2013}.
Therefore measurements of $n_{2,\parallel,e}$ and $n_{2,\perp,e}$ at $\theta = 0$ and $n_{2,\parallel,e}$ at $\theta = \pi/4$ are sufficient, in principle, to extract the three independent parameters.

\subsection{Raman contribution to $n_2$}
A well-understood but often neglected contribution to $n_2$ is the stimulated Raman effect \cite{levenson_interference_1972,levenson_dispersion_1974,wahlstrand_effect_2013,reichert_temporal_2014,wahlstrand_absolute_2015}.
To model this response, we apply an approach used previously for rotational and vibrational Raman contributions to $n_2$ in diatomic gases \cite{wahlstrand_effect_2013,wahlstrand_absolute_2015}.
The optical field $\bE(t)$ exerts a force on the lattice \cite{merlin_generating_1997}
\begin{equation*}
F_k(t) = \frac{\varepsilon_0}{2} \bE(t) \frac{\partial \chi}{\partial Q_k} \bE(t),
\end{equation*}
where $\partial \chi/\partial Q_k$ is the Raman susceptibility tensor for phonon mode $k$.
Modeling the phonon as a damped harmonic oscillator yields the coherent phonon amplitude
\begin{multline}
Q_k(t) = \\ \frac{\varepsilon_0}{2\Omega'} \int_{-\infty}^t e^{-\Gamma(t-t')/2}
\sin\left[\Omega'(t-t')\right]
\mathbf{E}^*(t') \cdot \frac{\partial \boldsymbol{\chi}}{\partial Q_k} \cdot \mathbf{E}(t') \, dt',
\label{Q_general}
\end{multline}
where $\Omega' = (\Omega^2-\Gamma^2/4)^{1/2}$, in which $\Omega$ is the phonon mode frequency and $\Gamma$ is the damping rate.
For our (001)-cut diamond sample at normal incidence, the optical fields are in the $xy$ plane, so the only relevant Raman tensor (for first-order Raman scattering) excites the $z$-polarized phonon,
\begin{equation*}
\frac{\partial \chi}{\partial Q_z} = R_z = \begin{bmatrix} 0 & d & 0 \\ d & 0 & 0 \\ 0 & 0 & 0\end{bmatrix},
\end{equation*}
where $d$ is a scalar characterizing the change in susceptibility with lattice displacement (see below).

Consider a Gaussian-shaped pump (excitation) field $\bE_e(t) = \bE_{e,0} \exp(-(t-t_d)^2/\tau_e^2) e^{-i\omega_e t} + \mathrm{c.c.}$, where $t_d$ is a time delay and $\tau_e$ characterizes the pulse duration.
If $\tau_e \lesssim \Omega^{-1}$, an oscillating coherent phonon response can be excited via ISRS \cite{merlin_generating_1997}.
In the opposite long-pulse limit ($\tau_e \gg \Omega^{-1}$), the phonon amplitude $Q$ adiabatically follows the magnitude squared of the pump field.
A closed-form solution to Eq.~(\ref{Q_general}) that applies for arbitrary $\tau_e$ is given in the Supplemental Material \cite{supp}.
For a pump beam linearly polarized at an angle $\theta$ with respect to the $[100]$ axis, the time-dependent phonon amplitude in the long pulse limit is
\begin{equation}
Q^s(t) = \frac{\varepsilon_0 d}{\Omega^2} |E_e(t)|^2 \sin(2\theta).
\end{equation}
In a pump-probe experiment, the optical properties are sampled by an optical probe field $\bE_p(t)$.
The phase and amplitude of the outwardly propagating probe field depend on $Q(t)$ because a polarization oscillating at the probe frequency is excited,
\begin{equation*}
\bP^s_R(t) = \varepsilon_0 \frac{\partial \chi}{\partial Q} Q^s(t) \bE_p(t).
\end{equation*}
We refer to this purely pump-driven contribution to $Q(t)$ and $\bP(t)$ as the ``smooth'' term (superscript s) because it comes from the pump irradiance profile, which is spatiotemporally smooth, unlike the ``grating'' term from pump-probe interference \cite{wahlstrand_effect_2013}.
In diamond in the long-pulse limit, projecting this polarization onto the probe polarization axis yields, for a probe polarized parallel to the pump (again, at $\theta$ with respect to $[100]$),
\begin{equation}
P^s_{\parallel,R}(t) = \frac{\varepsilon_0^2 d^2}{\Omega^2} |E_e(t)|^2 E_p(t) \sin^2(2\theta),
\label{PolQs_par}
\end{equation}
while for the probe polarized perpendicular to the pump,
\begin{equation}
P^s_{\perp,R}(t) = -\frac{\varepsilon_0^2 d^2}{\Omega^2} |E_e(t)|^2 E_p(t) \sin^2(2\theta).
\label{PolQs_perp}
\end{equation}

For an optical field affecting itself (i.e. self phase modulation), the smooth response is the entire effect.
In a pump-probe experiment, the probe response has an additional source.
Ignoring the Raman tensor (which imposes selection rules), the total field in the integrand of Eq.~(\ref{Q_general}) is $|E|^2 = |E_e + E_p|^2$, and for $|E_e| \gg |E_p|$, this can be approximated as $|E_e|^2 + (E_e^* E_p+c.c.)$.
The term in parentheses is the origin of the ``grating response'' (superscript g).
A closed-form solution to Eq.~(\ref{Q_general}) for the grating response for Gaussian pump and probe fields is given in the Supplemental Material \cite{supp}.
In the long-pulse limit, for parallel pump and probe fields polarized at an angle $\theta$ with respect to the [100] direction, the grating phonon amplitude is (for $\Gamma \ll \Omega$)
\begin{equation}
Q^g_\parallel(t) = \\ \frac{\varepsilon_0 d}{\Omega^2-\Delta\omega^2-i\Gamma\Delta\omega} E_e^*(t) E_p(t) \sin(2\theta) + \mathrm{c.c.},
\end{equation}
where $\Delta \omega = \omega_p - \omega_e$.
For perpendicular pump and probe fields,
\begin{equation*}
Q^g_\perp(t) = \\ \frac{\varepsilon_0 d}{\Omega^2-\Delta\omega^2-i\Gamma\Delta\omega} E_e^*(t) E_p(t) \cos(2\theta) +  \mathrm{c.c.}
\end{equation*}
Using
\begin{equation*}
\bP_R^g(t) = \varepsilon_0 \frac{\partial \chi}{\partial Q} Q^g(t) \bE_e(t),
\end{equation*}
the time-dependent phonon field combines with the pump field to produce a polarization in the probe direction,
\begin{align*}
P^g_{\parallel,R}(t) &= \frac{\varepsilon_0^2 d^2}{\Omega^2-\Delta\omega^2-i\Gamma\Delta\omega} |E_e(t)|^2 E_p(t) \sin^2(2\theta), \\
P^g_{\perp,R}(t) &= \frac{\varepsilon_0^2 d^2}{\Omega^2-\Delta\omega^2-i\Gamma\Delta\omega} |E_e(t)|^2 E_p(t) \cos^2(2\theta),
\end{align*}
to produce a field radiating in the probe direction.
Unlike the smooth term, this grating response depends sensitively on the frequency detuning $\Delta \omega$ between the pump and probe beams.

Each polarization expression is of the form $P(t) = \varepsilon_0 \Delta \chi_{\mathrm{eff}} E_p(t)$, where the effective susceptibility change $\Delta \chi_{\mathrm{eff}}$ is proportional to the pump irradiance, $I_e$.
We define an effective Kerr coefficient $n_{2,\mathrm{eff}} = \mathrm{Re} \Delta \chi_{\mathrm{eff}}/(2I_e n_0)$.
Summing the smooth and grating contributions and normalizing to $\n2x$ to define a Raman fraction $f_R = d^2/(4n_0^2 c \Omega^2 \n2x)$, we obtain
\begin{widetext}
\begin{align}
n_{2,\parallel,R}(\theta,\Delta\omega) &= \n2x f_R \mathrm{Re} \left[1 + \frac{\Omega^2}{\Omega^2-\Delta\omega^2-i\Gamma\Delta\omega} \right] \sin^2(2\theta), \label{Raman_par} \\
n_{2,\perp,R}(\theta,\Delta\omega) &= \n2x f_R \mathrm{Re} \left[ -\sin^2(2\theta) + \frac{\Omega^2}{\Omega^2-\Delta\omega^2-i\Gamma\Delta\omega} \cos^2(2 \theta) \right]. \label{Raman_perp}
\end{align}
\end{widetext}
In the expressions above, the first term in square brackets is the ``smooth'' contribution, while the second term is the ``grating'' contribution.
Comparing with Eqs.~(\ref{elec_par}) and (\ref{elec_perp}), we see that the smooth Raman term, which governs the response for a single beam, and thus would manifest in Z-scan as a pulse-duration-dependent $n_2$, has the same angle dependence as $\sigma$.
By contrast, the grating term, which contributes to the signal in a pump-probe experiment, exhibits a distinct $\cos^2(2\theta)$ angular dependence for pump and probe perpendicular along with a resonant dependence on the detuning $\Delta \omega$.

\section{Susceptibility measurement}
\begin{figure}
    \centering
    \includegraphics[width=8.5 cm]{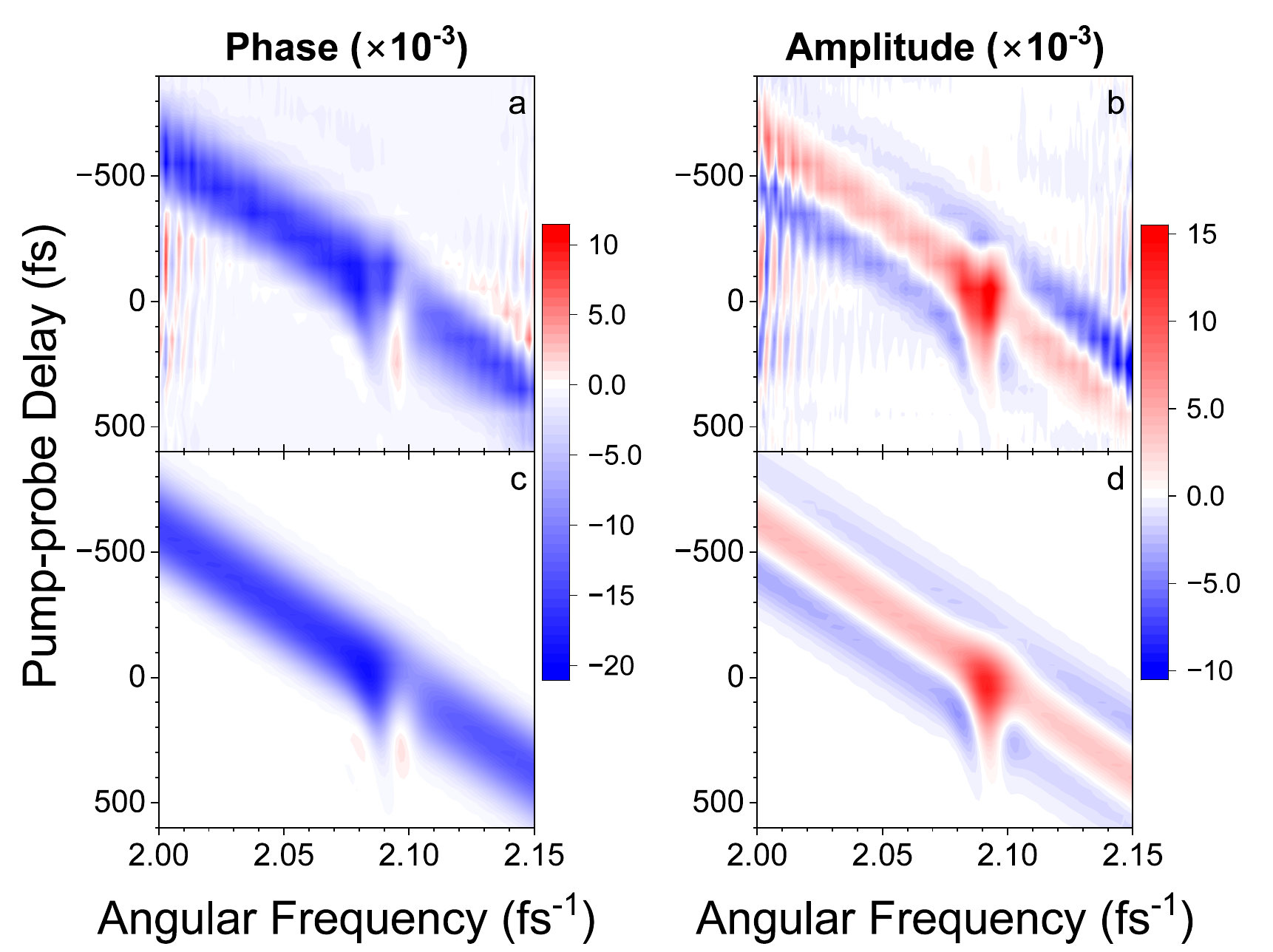}
    \caption{Pump-probe results in diamond and simulations for pump and probe beams polarized parallel to [110]. (a,b) Experimental results for 905~nm probe; (a) phase, (b) amplitude. (c,d) Simulations for 905~nm probe; (c) phase, (d) amplitude. The abrupt change in phase and amplitude at 2.09~fs$^{-1}$ is indicative of coherent vibrational excitation when the pump and probe are resonant with the diamond phonon mode.}
    \label{expt1}
\end{figure}

We use a phase-sensitive pump-probe technique based on spectral interferometry \cite{wahlstrand_absolute_2015,wahlstrand_precise_2026} to fully characterize the third-order nonlinear response of diamond near 1~$\mu$m wavelength.
Diamond's Raman spectrum is dominated by a single phonon mode at 1332~cm$^{-1}$ (39.9~THz) with a linewidth of approximately 2~cm$^{-1}$.
The pump pulse is derived from the 1024~nm output of a Yb-doped solid-state laser and has a full width at half maximum (FWHM) of 250~fs.
The probe pulse is the signal output of an optical parametric amplifier pumped by the same laser; in the experiments described here we tune its central wavelength between 860~nm and 940~nm.
At a given central wavelength its FWHM bandwidth is 18~nm, which corresponds to a transform-limited pulse duration of 70~fs.
We chirp the probe pulse using a ZnSe window so that its stretched duration exceeds the pump pulse duration, allowing single-shot capture of the full time-dependent response.
Near a probe wavelength of 901~nm, the phonon mode in diamond is resonantly excited.
The sample is a (001)-oriented crystal (Type IIa) grown by chemical vapor deposition (CVD).
The thickness of the sample is $L = (514 \pm 5)$~$\mu$m (all uncertainties are 1 standard deviation unless otherwise specified) determined via Fabry-Perot fringes in the transmission spectrum near 940~nm, using group refractive index $n_g = 2.42$ \cite{turri_index_2017}.
The sample is manufacturer-specified to have boron and nitrogen fractional concentrations of $<0.05 \times 10^{-6}$ and $<1 \times 10^{-6}$, respectively.
Figure \ref{expt1}ab shows the measured change in spectral amplitude and phase, for pump and probe parallel along the $[110]$ crystal direction, as a function of probe wavelength and pump-probe time delay.
From the dependence of the phase shift on pump-probe delay, we extract a probe group delay dispersion (GDD) of 3300~fs$^2$ \cite{wahlstrand_precise_2026}.
Figure \ref{expt1}cd shows simulations based on analytical integration of Eq.~(\ref{Q_general}) using $\Omega/(2\pi) = 39.9$~THz and $\Gamma = 0.4$~ps$^{-1}$, or $2$~cm$^{-1}$.

We measured $\n2x$ using a central probe wavelength of 926~nm with both the pump and probe fields along $[100]$.
For this polarization configuration ($\theta = 0$, parallel polarization), the Raman contribution is forbidden by selection rules.
The irradiance profile was characterized by sweeping the pump beam across the probe and measuring the full spatiotemporal response \cite{wahlstrand_precise_2026}.
Based on the linear dispersion of diamond \cite{turri_index_2017}, we estimate a temporal walkoff between pump and probe pulses of at most 14~fs, which is negligible for our 250~fs FWHM pump pulse duration.
At a peak pump irradiance of $(0.78 \pm 0.08)$~GW/cm$^2$, we measured a total cross-phase-modulation phase shift of 3.36~mrad with the diamond sample in place and a background phase shift of 200~$\mu$rad from the response of air.
Subtracting the air background yields $\n2x = (7.1 \pm 0.9) \times 10^{-16}$~cm$^2$/W, where we have divided the measured cross-phase-modulation phase shift by a factor of 2 to account for the two-beam coupling phase shift.
Unlike in our previous measurement in fused silica \cite{wahlstrand_precise_2026}, where the vibrational grating contribution decreased this factor, the complete absence of a Raman contribution in diamond for fields along $[100]$ ensures that the factor of 2 is exact.
Further details of the absolute $\n2x$ measurement, including uncertainty sources, can be found in the Supplemental Material \cite{supp}.

\begin{figure*}
    \centering
    \includegraphics[width=15cm]{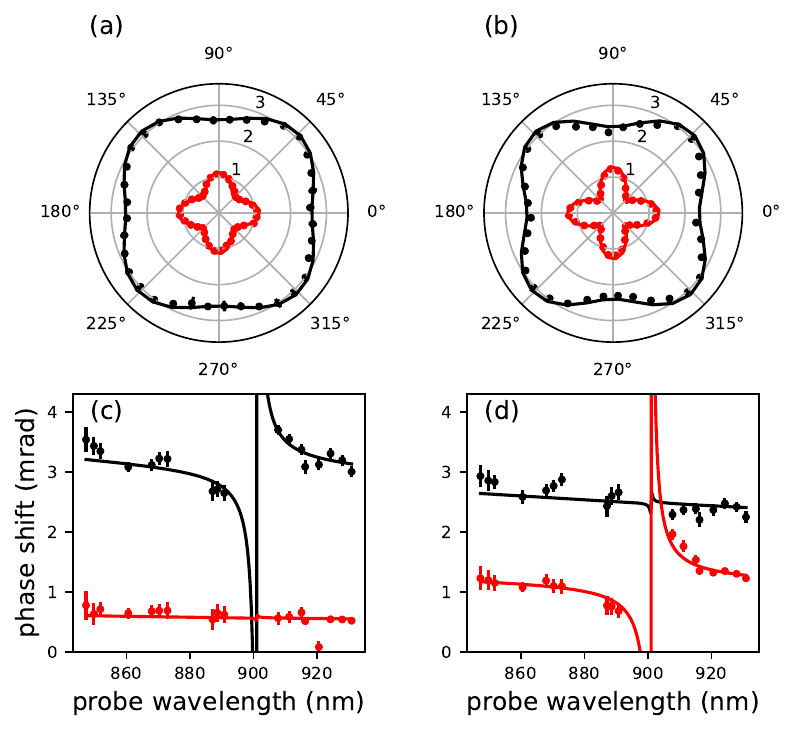}
    \caption{Phase shift versus sample angle ($\theta$) and probe wavelength. Points show measurements, and lines show fits to Eqs.~(\ref{elec_par},\ref{elec_perp},\ref{Raman_par},\ref{Raman_perp}). Black points and lines are for pump and probe polarization parallel, and red points and lines are for pump and probe perpendicular. (a,b) Sample-angle-dependent response for 1024~nm pump pulse at probe wavelengths (a) 860~nm and (b) 926~nm. (c,d) Probe-wavelength-dependent response for (c) $\theta = 45^\circ$ and (d) $\theta = 0^\circ$.}
    \label{angle}
\end{figure*}

To determine $\sigma$, $\delta$, and $f_R$, we measured the response for $\theta$ ranging from $0^\circ$ to $350^\circ$ in $10^\circ$ steps and for $\Delta \omega/(2\pi)$ ranging from 25~THz to 60~THz.
The phase shift as a function of probe frequency was extracted from the spectral interferometry measurements; we excluded points too close to the phonon resonance at $\Delta \omega = \Omega$ because the signal there is distorted from the long-pulse-limit approximation [Eqs.\ (\ref{elec_par},\ref{elec_perp},\ref{Raman_par},\ref{Raman_perp})].
The approximately 12~THz usable width of the probe spectrum allowed multiple phase-shift measurements for a given probe central wavelength.
To cover the full range of $\Delta \omega/(2\pi)$, we acquired data using probe central wavelengths of 926~nm, 905~nm, and 860~nm.
Figure~\ref{angle} shows representative slices of the resulting three-dimensional dataset: the measured phase shifts and their uncertainties (calculated from the standard deviation of multiple measurements) are plotted as points with error bars, while the solid curves show a global weighted-least-squares fit to Eqs.\ (\ref{elec_par},\ref{elec_perp},\ref{Raman_par},\ref{Raman_perp}) with $\Omega/(2\pi) = 39.9$~THz and $\Gamma = 0.4$~ps$^{-1}$ held fixed.
Because the values of $\Delta \omega$ used in the fits are relatively far from the resonance $\Omega$, the exact value chosen for $\Gamma$ does not significantly affect the fit parameters.
We find $\sigma = -0.39 \pm 0.04$, $\delta = 0.28 \pm 0.03$, and $f_R = 0.063 \pm 0.006$.
Uncertainty estimates for these parameters are calculated from the fit using weighted-least-squares standard errors.
We have repeated the fit with an expression derived from a more general analysis that accounts for the finite duration of the pulses; the result is the same within uncertainty so we do not include it here.
These measurements, combined with the value of $\n2x$ mentioned previously, constitute a full characterization of the third-order nonlinear response near 1~$\mu$m wavelength.

\section{Discussion}

Our measured parameters are summarized in Table \ref{data} alongside results from the literature.
Where multiple wavelengths were used, we show only the value for the nearest wavelength to 1~$\mu$m; where multiple types of samples were measured, we show only the value for the sample with the smallest density of impurities.
It should be kept in mind that both Motojima \emph{et al.}\ \cite{motojima_giant_2019} and Talik \emph{et al.}\ \cite{talik_impact_2025} reported a dependence of $n_2$ on impurity density, so variations in sample purity may be responsible for some of the scatter in reported $n_2$ values.
Also notably, except for Levenson and Bloembergen \cite{levenson_dispersion_1974} and Talik \emph{et al.} \cite{talik_impact_2025}, the previous studies to which we compare our results did not uniquely specify the laser polarization with respect to the crystal axes.
It is common practice to compare $n_2$ measurements to the two-parabolic-band model of Sheik-Bahae \emph{et al.}\ \cite{sheik-bahae_dispersion_1991}, $n_2 \approx K' E_p^{1/2}/(n_0 E_g^4) G_2(\hbar\omega/E_g)$, where $E_p = 21$~eV, $K' = 0.94 \times 10^{-8}$ (yielding $n_2$ in esu for energies in eV), and $G_2(x)$ is a dispersion function for two-photon absorption.
Using the direct bandgap $E_g = 7.2$~eV \cite{logothetidis_origin_1992} yields $n_2 \approx 2.1 \times 10^{-16}$~cm$^2$/W.
Using the indirect bandgap $E_{g,i} = 5.5$~eV gives $n_2 \approx 6.8 \times 10^{-16}$~cm$^2$/W, which is in closer agreement with our experimental value.

\begin{table*}
\centering
\begin{ruledtabular}
\begin{tabular}{l | c c c c c}
 & $\lambda$ (nm) & $\n2x$ (cm$^2$/W) & $\sigma$ & $\delta$ & $f_R$ \\
\hline
This work & 975\footnote{Effective wavelength calculated as mean of pump and probe wavelengths.} & $(7.1 \pm 0.9) \times 10^{-16}$ & $-0.39 \pm 0.04$ & $0.28 \pm 0.03$ & $0.063 \pm 0.006$\\
Levenson \emph{et al.}\footnote{Three-wave mixing, referenced to previous measurement of Raman polarizability. We multiply by 4 to account for the differing convention used.} \cite{levenson_dispersion_1974} & 545 & $1.24 \times 10^{-15}$ & $-0.15 \pm 0.06$ & $0.33 \pm 0.02$ & 0.061 \\
Sheik-Bahae \emph{et al.}\footnote{Z-scan.} \cite{sheik-bahae_optical_1995} & 1064 & $(4.2 \pm 0.6) \times 10^{-16}$\footnote{Light polarization with respect to crystal orientation not specified.} & --- & --- & --- \\
Koz\'ak \emph{et al.}\footnotemark[3] \cite{kozak_two-_2012} & 636 & $5.5 \times 10^{-16}$\footnotemark[4] & --- & --- & --- \\
Hausmann \emph{et al.}\footnote{Measured threshold for parametric oscillation in a ring resonator device.} \cite{hausmann_diamond_2014} & 1600 & $(8.2 \pm 3.5) \times 10^{-16}$\footnotemark[4] & --- & --- & --- \\
Almeida \emph{et al.}\footnotemark[3] \cite{almeida_nonlinear_2017} & 955 & $(9.5 \pm 1.4) \times 10^{-16}$\footnote{Laser polarized perpendicular to the $[100]$ axis.} & --- & --- & --- \\
Motojima \emph{et al.}\footnotemark[3] \cite{motojima_giant_2019} & 800 & $4.2 \times 10^{-16}$\footnotemark[4] & --- & --- & --- \\
Talik \emph{et al.}\footnotemark[3] \cite{talik_impact_2025} & 1032 & $3.155 \times 10^{-16}$ & $0.454 \pm 0.057$ & --- & --- 
\end{tabular}
\end{ruledtabular}
\caption{Summary of our measured parameters and comparison with the literature.}
\label{data}
\end{table*}

We note that our measurement of $\sigma$ agrees in magnitude with that of Talik \emph{et al.}\ \cite{talik_impact_2025}, but is opposite in sign.
Our finding of $\sigma <0$ is typical for cubic crystals at photon energies well below the band gap.
In regimes where $n_2$ is dominated by virtual two-photon transitions, as expected here, Hutchings and Wherrett \cite{hutchings_theory_1995}, building on prior work \cite{dvorak_measurement_1994}, argued that $\sigma \approx -2E_g/(E_g+E_u)$, where $E_g$ is the fundamental band gap and $E_g+E_u$ is the energy separation between the valence band and the next-higher conduction band.
While this simple formula works reasonably well for GaAs and InSb \cite{hutchings_theory_1995},
applying it to diamond using the direct bandgap $E_g = 7.2$~eV \cite{logothetidis_origin_1992} and $E_g + E_u = 15$~eV \cite{hybertsen_electron_1986} yields $\sigma = -0.96$, far higher in magnitude than our measurement.
Finally, our measured $\sigma$ and $\delta$ values roughly satisfy Kleinman symmetry, $\delta \approx (2+\sigma)/6$, as expected for photon energies far from any resonances.

The strength of Raman scattering by a phonon mode is often characterized by a Raman polarizability $\alpha_R = (\mu/N_{\mathrm{cell}})^{1/2} d/(4\pi)$, where $\mu$ is the reduced mass of the phonon mode and $N_{\mathrm{cell}}$ is the number of unit cells per unit volume (in diamond, $\mu = 9.97 \times 10^{-24}$~g and $N_{\mathrm{cell}}= 8.81 \times 10^{22}$~cm$^{-3}$), and the factor $4\pi$ converts to the Gaussian-unit susceptibility used in the literature definition \cite{cardona_resonance_1982}.
Note that $\alpha_R$ is generally given in units of area.
We can combine our absolute measurement of $\n2x$ with our measurement of $f_R$ to find $|d| = 2 n_0 \Omega (c \n2x f_R)^{1/2} = (1.39 \pm 0.11) \times 10^9$~kg$^{-1/2}$ m$^{1/2}$, and $|\alphaR| = (n_0\Omega/2\pi)(\mu c
  \n2x f_R/N_{\mathrm{cell}})^{1/2} = (3.7 \pm 0.3) \times 10^{-16}$~cm$^2$.
The phonon in diamond is one of the most-studied Raman scatterers, with Cardona providing a recommended value at 514.5~nm of $|\alpha_R| = (4.3 \pm 0.6) \times 10^{-16}$~cm$^2$ \cite{cardona_resonance_1982} largely based on the measurement of Grimsditch and Ramdas, who used Raman scattering cross-section measurements referenced to Brillouin scattering \cite{grimsditch_brillouin_1975}.
To account for the dependence of the Raman polarizability on photon energy $E$, Cardona proposed $\alpha_R(E) \propto g(E/E_g)$ \cite{cardona_resonance_1982}, where $g(x) = x^{-2} [2-(1-x)^{-1/2} - (1+x)^{-1/2}]$ is a dispersion formula developed by Calleja \emph{et al.} \cite{calleja_resonant_1978}.
If we translate the recommended value at 514.5~nm to our effective wavelength of 975~nm using this formula, we find $|\alpha_R| = (3.9 \pm 0.5)\times 10^{-16}$~cm$^2$, which is in agreement with our measured value.
However, the commonly accepted formula is $|\alpha_R(E)| = (6.5 \times 10^{-16}~\mathrm{cm}^2)g(E/E_g)$ \cite{mildren_optical_2013}, which is about 30~\% higher than Cardona's recommended value at 514.5~nm.
Using spontaneous Raman scattering at 1064~nm, Aggarwal \emph{et al.} measured $|\alphaR| = (8.1 \pm 0.8) \times 10^{-16}$~cm$^{2}$ \cite{aggarwal_measurement_2012}.
The density functional theory (DFT) calculation reported by Aggarwal \emph{et al.}~of $|\alphaR| = 5.0 \times 10^{-16}$~cm$^{2}$ at 1064~nm \cite{aggarwal_measurement_2012} is closer to the commonly accepted value.
Finally, in the previous section we mentioned that variations in impurity density have been found to affect $n_2$.
Based on their theoretical calculations, Aggarwal \emph{et al.} concluded that the presence of impurities does not strongly affect the Raman polarizability \cite{aggarwal_measurement_2012}, but we cannot rule out the possibility that localized microscopic strain, which is known to occur in CVD-grown diamond \cite{pinto_theory_2009,hoa_birefringence_2014}, may affect our measurement because of induced birefringence.

The most comprehensive prior measurement of the third-order susceptibility in diamond was performed by Levenson and Bloembergen in 1974 \cite{levenson_dispersion_1974}.
They employed wavelength-dependent three-wave mixing referenced to earlier measurements of $|\alphaR|$ \cite{mcquillan_stimulated_1970,gorelik_1970,anastassakis_electric-field-induced_1970}, which they scaled across wavelength using a phenomenological dispersion formula.
From their scaled reference value of $|\alphaR| = 4.9 \times 10^{-16}$~cm$^2$ and reported $\chi^{(3)}$ at 545~nm, we calculate $\n2x = 1.24 \times 10^{-15}$~cm$^2$/W.
Re-evaluating their 545~nm result using our measured $|\alpha_R|$ (scaled by wavelength using the Cardona formula) reduces their inferred value to $\n2x = 8.5 \times 10^{-16}$~cm$^2$/W; further scaling that value from 545~nm to our effective wavelength of 975~nm using the Sheik-Bahae dispersion formula, we find $\n2x = 5.9 \times 10^{-16}$~cm$^2$/W, which is not far from our measured value.
Using polarization-dependent measurements, Levenson and Bloembergen also determined each electronic $\chi^{(3)}$ tensor element \cite{levenson_dispersion_1974}, from which anisotropy parameters $\sigma = -0.15 \pm 0.06$ and $\delta = 0.33 \pm 0.02$ were calculated.
The difference between our measured values of $\sigma$ and $\delta$ and those of Levenson and Bloembergen is not surprising, given the very different wavelengths measured (their 545~nm versus our mean wavelength of 975~nm) and the fact that the anisotropy parameters can vary strongly with wavelength \cite{hutchings_theory_1995}.
Furthermore, using their 545~nm data, we calculate $f_R = 16\pi^2 \varepsilon_0 N_{\mathrm{cell}}|\alphaR|^2/(3 \mu \Omega^2 \chi_{xxxx}) = 0.061$, similar to our measured value.

As in the Levenson and Bloembergen experiment \cite{levenson_dispersion_1974}, our data includes spectroscopically distinct contributions from the Raman effect and the electronic $\chi^{(3)}$, and this provides internal consistency checks.
First, the Raman selection rules confirm the crystal orientation with respect to the light polarization.
Second, our measurement of the Raman polarizability provides an independent check on our peak irradiance calibration, which is often the largest source of systematic uncertainty in absolute $n_2$ measurements.
Specifically, if we overestimated our peak irradiance by a factor of 2, then our $n_2$ measurement would fall by a factor of 2 and would be in agreement with Sheik-Bahae \emph{et al.} \cite{sheik-bahae_optical_1995}, Talik \emph{et al.} \cite{talik_impact_2025}, and Motojima \emph{et al.} \cite{motojima_giant_2019}.
But if that were the case, then our measurement of $|\alphaR|$ would drop by a factor of $\sqrt{2}$, placing it well outside the range of previous measurements.
We are thus confident in our measurement of $n_2$ in this sample; deciding whether the wide range of $n_2$ measurements in diamond is caused by variation from sample to sample or the general difficulty of the measurement will require further effort.

\begin{figure}
    \centering
    \includegraphics[width=8.5 cm]{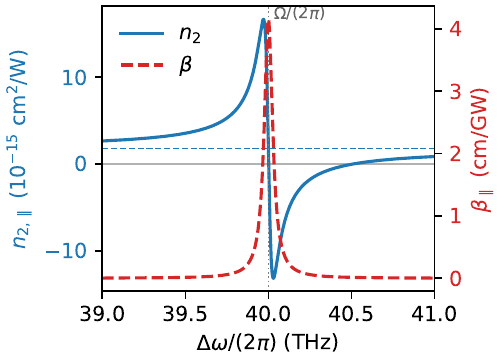}
    \caption{Total nonlinear response for pump and probe parallel to $\langle 110 \rangle$ for $\Delta \omega$ near the phonon resonance. The blue (red) curve shows the effective Kerr (two-photon absorption) coefficient as a function of $\Delta \omega$. The horizontal blue dashed line shows $n_{2,\parallel}$ for $\Delta \omega = 0$.}
    \label{resonance}
\end{figure}

Due to the resonant grating response from the Raman contribution to $n_2$, the nondegenerate nonlinear coefficients are strongly enhanced when $\Delta \omega \approx \Omega$, as shown in Fig.~\ref{resonance}.
Using Eqs.\ (\ref{elec_par}) and (\ref{Raman_par}) for a pump wavelength of 1024~nm and both pump and probe polarizations along $[110]$, $n_2$ in the long-pulse limit can be as high as $15\times 10^{-15}$~cm$^2$/W just below the resonance and as low as $-12\times 10^{-15}$~cm$^2$/W just above the resonance.
At these near-resonant frequencies, the effective anisotropy parameter $\sigma$ likewise becomes extreme.
However, this near-resonant enhancement is accompanied by strong nonlinear loss, specifically through Raman-mediated two-photon absorption.

\section{Two-photon Raman absorption}
Raman-mediated two-photon absorption is the operational mechanism behind stimulated Raman photothermal (SRP) microscopy \cite{zhu_stimulated_2023}.
We can describe the effective two-photon absorption coefficient near $\Delta \omega = \Omega$ as 
\begin{widetext}
\begin{align}
\beta_{\parallel,R}(\theta,\Delta \omega) &= \n2x \frac{4\pi}{\lambda_p} f_R \mathrm{Im} \left[ \frac{\Omega^2}{\Omega^2-\Delta\omega^2-i\Gamma\Delta\omega} \right] \sin^2(2\theta), \label{Raman_beta_par} \\
\beta_{\perp,R}(\theta,\Delta \omega) &= \n2x \frac{4\pi}{\lambda_p} f_R \mathrm{Im} \left[ \frac{\Omega^2}{\Omega^2-\Delta\omega^2-i\Gamma\Delta\omega} \right] \cos^2(2 \theta), \label{Raman_beta_perp}
\end{align}
\end{widetext}
where $\lambda_p$ is the wavelength of the probe. In diamond, for co-polarized pump and probe along $[110]$, or cross-polarized fields with one along $[100]$ and the other along $[010]$, the effective two-photon absorption coefficient can be as high as 4~cm/GW on resonance for our assumed value of $\Gamma$, as shown in Fig.~\ref{resonance} in red.
The effective two-photon absorption coefficient $\beta$ is related to the Raman gain coefficient $g_R$ at Stokes frequency $\omega_S$ via $g_R = \omega_S \beta/(\omega_S+\Omega)$, and the value of $g_R = 3.4$~cm/GW deduced from our measurement of $f_R$ and $\n2x$ is low compared to commonly used values in the literature \cite{mildren_highly_2009}, though both $\beta$ and $g_R$ scale with $\Gamma^{-1}$.

In addition to modifying the probe beam, the coherent interaction of the pump and probe pulses increases the population in the phonon modes, depositing heat.
This resonant process is the basis for the recently developed stimulated Raman photothermal (SRP) microscopy technique \cite{zhu_stimulated_2023,zhu_stimulated_2026} but has not been reported in solids, to our knowledge.
Under the ideal long-pulse conditions where Eqs.~(\ref{Raman_beta_par},\ref{Raman_beta_perp}) apply, the thermal energy deposited per unit volume $U$ is given by the work done by the laser on the phonon mode,
\begin{math}
U = \int F(t) \dot{Q}(t) dt.
\end{math}
As derived in the Supplemental Material \cite{supp} (where the general case of arbitrarily chirped pulses is also treated), for long pulses with matched chirp rates the instantaneous volumetric heating rate is
\begin{equation}
\dot{U}(t) = \frac{\Delta\omega}{\omega_p}\,\beta_R(\theta,\Delta\omega)\, I_e(t)\, I_p(t).
\label{Udot_cw}
\end{equation}
To investigate this two-photon Raman heating effect in solids, which could have advantages for thermal conductivity metrology as it avoids electronic excitation, we performed a second experiment.
To resonantly excite the sample, we use chirped pulses centered at 1030~nm, generated by a Yb laser, and a tunable pulse centered between 895~nm and 915~nm, generated by an OPA.
A different Yb laser and OPA were used as they produce narrower spectra that are nearly matched in bandwidth.
Both excitation pulses are chirped using grating-based stretchers that are tuned carefully such that the pulses have matching group delay dispersion.
This chirp reduces the peak irradiance by a factor of 40 from the transform-limited values, which allows us to increase the pulse fluence well beyond the damage threshold for transform-limited pulses. The pulse durations of the 1030~nm and 905~nm pump pulses are measured to be 11~ps and 15~ps respectively using a cross correlation with an unchirped 1030~nm pulse.

We directly measured the energy deposited via resonant Raman absorption by comparing the total transmitted energy of the excitation pulses through the diamond sample when they were overlapped in space and time to when they were temporally separated by 80~ps.
With incident pulse energies of 4~$\mu$J at 1030~nm and 1~$\mu$J at 905~nm, the total transmitted energy (about 3.4~$\mu$J after accounting for Fresnel losses) decreased by 44.6~nJ when overlapped in time, corresponding to energy deposited into optical phonons.
When the higher-frequency beam was detuned from 905~nm to either 915~nm or 895~nm ($\Delta \omega/(2\pi c) \approx \pm 122$~cm$^{-1}$ from the $\Omega/(2\pi c)=1332$~cm$^{-1}$ resonance) the transmitted energy loss dropped by more than a factor of 7.
We estimate an uncertainty in the measured energy deposition of about a factor of 2.
The absence of measurable energy loss under detuned excitation rules out significant electronic excitation or other nonresonant two-beam loss mechanisms, confirming that the depleted energy is transferred to the lattice via resonant Raman absorption. As a further check, we measured the photon loss in the 905~nm beam and the corresponding photon gain in the 1030~nm beam. As expected, the photon loss equaled the photon gain within experimental uncertainty.
Given the 47~$\mu$m FWHM beam diameters and a confocal parameter exceeding the sample thickness, we calculate a peak deposited energy density of 100~mJ/cm$^3$, where the effective spatial width is reduced by $\sqrt{2}$ to account for the quadratic dependence on the irradiance (product of the two profiles).
Dividing by $\hbar\Omega \approx 165$~meV yields a peak excited phonon density of $3.8 \times 10^{18}$~cm$^{-3}$ in the 1332~cm$^{-1}$ mode.
Compared to the primitive unit-cell density of $8.81 \times 10^{22}$~cm$^{-3}$ in diamond, this corresponds to a fractional mode occupancy of $4.3 \times 10^{-5}$ per unit cell ($2.1 \times 10^{-5}$ per atom), nearly two orders of magnitude smaller than the room-temperature thermal occupancy of $1.6 \times 10^{-3}$.
Following relaxation, the resulting peak temperature rise is approximately 58~mK.

The peak deposited energy density of 100~mJ/cm$^3$ may be compared to the prediction of Eq.~(\ref{Udot_cw}) and the more general expression for arbitrarily chirped pulses derived in the Supplemental Material \cite{supp}.
Accounting for the measured beam waist, spectral bandwidth (roughly 35~\% broader for the 905~nm pulse), pulse duration, and the 17~\% reflection loss at the entrance face, we calculate an irradiance of $I_{\mathrm{1030\,nm}} = 14.3$~GW/cm$^2$ and $I_{\mathrm{905\,nm}} = 2.6$~GW/cm$^2$.
For perfect chirp matching, we calculate 448~mJ/cm$^3$, about a factor of 4.5 higher than our measured value.
However, the amount of heat deposited is highly sensitive to pulse chirp: a 12~\% mismatch in group delay dispersion alone is sufficient to reduce the predicted energy density to the observed value of 100~mJ/cm$^3$.
We attribute the difference between the ideal prediction and our measurement to a combination of chirp mismatch (including finite third-order chirp) and imperfect spatial overlap of the two beams.

To observe the time-dependent temperature evolution in the diamond, we monitored the thermal response via photothermal beam deflection \cite{jackson_photothermal_1981} using a continuous-wave HeNe laser ($\lambda = 633$~nm).
In this technique, a probe beam passes through the excited sample and its direction in the far field is modified by the refractive index gradient from the temperature profile produced by the excitation pulses.
The beam deflection is measured using a segmented detector, and the signal is proportional to $dn/dx$ (the spatial gradient of the refractive index), which is proportional to $dT/dx$, the laser-induced thermal gradient.
To relate the beam deflection experiment to the temperature rise, we solve the heat flow equation, $\alpha \nabla^2 T = dT/dt$ assuming a cylindrically symmetric source, where $\alpha$ is the thermal diffusivity of diamond.
The initial temperature distribution immediately following laser excitation is modeled as
\begin{equation}
T(x,y,t=0) = T_0 \exp \left(-\frac{x^2+y^2}{R_0^2} \right) + T_a,
\end{equation}
where $R_0$ is the initial size of the thermal profile, $T_0$ is the peak temperature rise due to absorbed laser energy, and $T_a$ is the ambient temperature.

\begin{figure}
    \centering
    \includegraphics[width=8.5 cm]{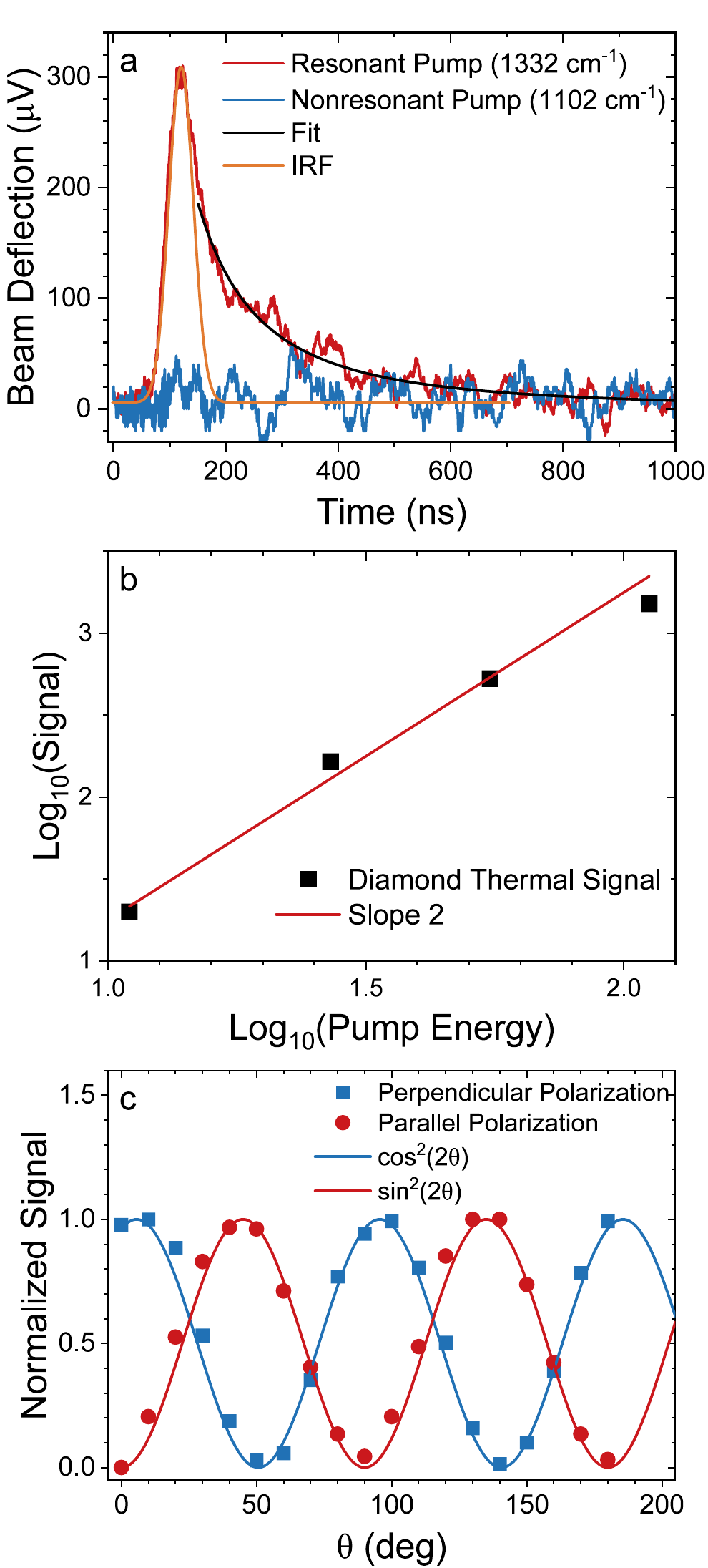}
    \caption{Beam deflection results. (a) HeNe probe beam deflection as a function of time following excitation by a two-color pump pulse. (b) Dependence of beam deflection signal on total pump pulse energy with a line of slope 2 (expected for two-photon absorption). (c) Polarization dependence of the beam deflection signal along with the scaled prediction of Eqs.~(\ref{Raman_beta_par}) and (\ref{Raman_beta_perp}).}
    \label{heat}
\end{figure}

The beam deflection is proportional to the transverse temperature gradient $dT/dx$ evaluated at the probe beam offset $x$, given by
\begin{equation}
\frac{dT}{dx}(x,y=0,t) = -2xT_0 \frac{R_0^2}{(R_0^2+4\alpha t)^2} \exp \left(-\frac{x^2}{R_0^2+4\alpha t} \right).
\label{BeamDef}
\end{equation}
The measured time-dependent beam deflection after resonant two-photon Raman absorption is shown as a red curve in Fig.~\ref{heat}a while the nonresonant response is shown in blue.
The absence of a measurable signal under detuned excitation verifies that electronic excitation or other nonlinear two-beam effects are not the source of the beam deflection.
Instead, phonons are generated after Raman absorption leading to a change in temperature that causes the probe beam to deflect as a function of time.
Evaluating Eq.~(\ref{BeamDef}) using the experimental parameters ($R_0 = 24$~$\mu$m, $x = 12$~$\mu$m, and $\alpha = 11$~cm$^2$/s; black curve) yields reasonable agreement with the measured transient.
The orange curve is the instrument response function (IRF) of the detector to a femtosecond laser pulse, showing the limited bandwidth of the segmented detector.
The experimentally measured beam deflection is consistent with a pump-induced temperature rise and subsequent thermal diffusion as described by Eq.~(\ref{BeamDef}).
Diamond is a particularly challenging material to measure because of its extremely high value of $\alpha$ (and thus rapid heat transfer out of the probed region) and low value of thermo-optic coefficient $dn/dT$ (and thus small beam deflection).

To measure the dependence of the deflection signal on excitation pulse energy, relative polarization, and crystal orientation ($\theta$), we inserted an optical chopper into the excitation path and recorded the signal using a lock-in amplifier.
Figure \ref{heat}b shows the energy dependence of the beam-deflection signal under resonant excitation on a log-log plot; the observed quadratic dependence confirms that the beam-deflection signal is the result of two-photon absorption.
The polarization dependence of the thermal signal is shown in Fig.~\ref{heat}c.
For pump and probe beams with parallel polarization, we find that the signal is highest for fields along the $[110]$ direction.
For pump and probe beams with perpendicular polarization, we find that the signal is highest for pump and probe fields along the $[100]$ and $[010]$ axes, respectively.
These observations are consistent with Eqs.~(\ref{Raman_beta_par}) and (\ref{Raman_beta_perp}).

Altogether, these results demonstrate that when the pump lasers are resonant with the 1332~cm$^{-1}$ diamond phonon mode, energy is deposited into phonons that relax rapidly generating heat, which subsequently diffuses, in agreement with Eq.\ (\ref{BeamDef}).
For thermal conductivity metrology, resonant Raman absorption offers a route to material-specific, localized heating without generating free carriers.
Because free-carrier dynamics can obscure pump-probe-based measurements of thermal transport \cite{opsal_thermal_1985,wang_thermal_2016}, sub-bandgap resonant Raman excitation may provide distinct advantages over above-bandgap or multiphoton electronic excitation.
This resonant heating mechanism is also directly relevant to high-power diamond Raman lasers, where the pump and Stokes fields are, by construction, separated in frequency by $\Omega$ \cite{mildren_highly_2009}.
Diamond's high thermal conductivity has enabled kilowatt-class output powers with low thermal lensing \cite{antipov_12_2019}, but residual heating remains a key limitation to further power scaling \cite{bai_comprehensive_2021}.
Thermal modeling has generally treated phonon-generation heating phenomenologically, without isolating its dependence on pump-Stokes detuning or polarization.
The Raman-mediated two-photon absorption coefficient calculated here from our measurements of $\n2x$ and $f_R$ supplies a quantitative heat-source term, distinct from linear defect absorption and from free-carrier absorption via electronic two-photon transitions, that could be incorporated directly into such thermal models.

\section{Conclusions}
In summary, we have fully characterized the third-order nonlinear optical response of diamond near 1~$\mu$m wavelength, explicitly resolving the bound electronic tensor components and the stimulated Raman contribution.
In the long-pulse limit for co-polarized light along $\langle 110\rangle$, the Raman contribution accounts for a few percent of the total effective $\chi^{(3)}$ response.
Under two-color nondegenerate excitation, tuning the frequency difference near the 40~THz optical phonon resonance dramatically modifies the effective $n_2$ and induces an extremely anisotropic response.
Accompanying this near-resonant enhancement is strong Raman-mediated two-photon absorption.
Finally, using dual chirp-matched excitation pulses in combination with photothermal beam deflection, we directly observe localized lattice heating driven by this resonant Raman absorption, establishing a quantitative, polarization-resolved heat-source term for sub-bandgap thermal metrology and high-power Raman laser modeling.

The technique demonstrated here for characterizing $\chi^{(3)}$ can readily be applied to other solid-state materials; however, probing the contribution of modes with smaller phonon frequencies $\Omega$ will necessitate improved pump-light rejection.
In our current configuration, measuring $\Delta \omega/(2\pi c)$ down to 300~cm$^{-1}$ is feasible, and refined filtering should push this limit lower.
For modes below this range, probing the Raman response via impulsive stimulated Raman scattering (ISRS) with ultrashort pulses remains straightforward \cite{smolorz_measurement_1999}.
Combining nondegenerate pump-probe spectroscopy with ISRS thus provides a comprehensive pathway toward full characterization of the Raman contribution to $\chi^{(3)}$ across all frequency regimes.
In non-centrosymmetric crystals containing modes that are both Raman- and infrared-active, phonon-polariton dispersion must also be accounted for.
Crucially, the Raman contribution governing the single-beam $n_2$ involves near-forward scattering, whereas Raman spectra found in the literature are typically captured via backscattering or 90$^\circ$ scattering and therefore probe phonons with larger wavevector.
For example, in ZnSe, the dominant excited phonon-polariton frequency shifts with excitation wavelength due to group-velocity-matching constraints \cite{stevens_cherenkov_2001}.
Finally, in absorbing materials where electronically resonant Raman scattering occurs \cite{stevens_coherent_2002}, phase-sensitive measurements in reflection mode \cite{cruz_phase-sensitive_2024} could quantify the contribution of resonant Raman contributions to both $n_2$ and two-photon absorption, potentially providing a valuable benchmark for first-principles theoretical calculations.


\begin{acknowledgments}
We thank Adam Biacchi for useful advice regarding the sample and Andrea Centrone for valuable discussions. J.K.W. and C.D.C. acknowledge support from the NIST CHIPS Metrology program.
\end{acknowledgments}






%

\end{document}


\title{Supplemental document for \\ ``Raman-induced Kerr effect and heating in diamond''}
\author{C. D. Cruz}
\affiliation{Nanoscale Device Characterization Division, National Institute of Standards and Technology, Gaithersburg, MD 20899 USA}
\author{J. C. Stephenson}
\affiliation{Associate, Sensor Science Division, National Institute of Standards and Technology, Gaithersburg, MD 20899 USA}
\author{J. K. Wahlstrand}
\affiliation{Nanoscale Device Characterization Division, National Institute of Standards and Technology, Gaithersburg, MD 20899 USA}
\date{\today}

\maketitle

\def\n2x{n_{2,\langle 100 \rangle}}
\def\bE{\mathbf{E}}\def\bEe{\mathbf{E}_e}\def\bEp{\mathbf{E}_p}
\def\br{\mathbf{r}}
\def\bk{\mathbf{k}}\def\bke{\mathbf{k}_e}\def\bkp{\mathbf{k}_p}
\def\Fourier{\mathcal{F}}
\newcommand{\chiph}{\chi_{\mathrm{ph}}}
\newcommand{\dd}{\mathrm{d}}

\maketitle

\section{Theory of Raman response}

If we assume Gaussian pump and probe pulses (allowing the probe pulse to be chirped),
\begin{eqnarray}
E_e(t) &=& E_{e,0} \exp(-t^2/\tau_e^2) e^{-i\omega_e t} + c.c., \\
E_q(t) &=& E_{q,0}\frac{\tau_0}{\sqrt{\tau_0^2-2iG}} \exp(-t^2/\tau_q^2) \exp\left(-i\omega_q t - i\xi t^2\right)+c.c.,
\label{Gaussian_pulse}
\end{eqnarray}
where $\tau_q^2 = \tau_0^2+4 G^2/\tau_0^2$ and $\xi = 2G/(\tau_0^2 \tau_q^2)$, in which $G$ is the group delay dispersion (with units of time squared, positive for normal dispersion), and $\tau_0$ is the transform-limited pulse duration, we can derive closed form solutions to Eqs.~(4,7) from the main text.
The solution for the smooth term [Eq.~(4)] is
\begin{equation}
Q_s(t) = \frac{\varepsilon_0}{\Omega'} \frac{\partial \chi}{\partial Q} |E_{e,0}|^2\, \frac{\sqrt{\pi}}{2\sqrt{\eta}}\, e^{-\eta(t-t_d)^2}\, \mathrm{Im}\left[\mathrm{erfcx}(z)\right],
\label{Qvib_smooth_pulse_closed}
\end{equation}
where $\eta = 2/\tau_e^2$ and $z = (\Gamma/2 - i\Omega')/(2\sqrt{\eta}) - \sqrt{\eta}(t-t_d)$.
The function $\mathrm{erfcx}(z)$ is the scaled complementary error function.

The solution for the grating term [Eq.~(7)] is, defining $\Delta\omega = \omega_q - \omega_e$,
\begin{equation}
Q_g(t) = -\frac{\varepsilon_0 E_{e,0}^*E_{q,0}}{\Omega'}\frac{\partial\chi}{\partial Q}\frac{i}{2}\,e^D\,\frac{\sqrt\pi}{2\sqrt C}\Big[\mathrm{erfcx}(z_+)-\mathrm{erfcx}(z_-)\Big] + c.c.,
\label{Qvib_grating_pulse_closed}
\end{equation}
with $z_\pm \equiv -\frac{B_0-\Gamma/2 \pm i\Omega'}{2\sqrt{C}}$, $C \equiv \frac{1}{\tau_e^2}+\frac{1}{\tau_q^2}+i\xi$,
\begin{equation}
B_0 \equiv \frac{2(t-t_d)}{\tau_e^2}+\frac{2t}{\tau_q^2}+2i\xi t - i\Delta\omega, \quad D \equiv -\frac{(t-t_d)^2}{\tau_e^2}-\frac{t^2}{\tau_q^2}-i\xi t^2+i\Delta\omega t,
\end{equation}
and where the principal branch of $\sqrt{C}$ (positive real part) is used.

These were used in the expressions for $P(t)$ and summed to generate Figs.~1cd.
The amplitude of the simulation was scaled to match the experiment (this data set did not have carefully characterized irradiance and pump-probe overlap, so the absolute magnitude of the phase shift and amplitude change are not meaningful).

\section{Raster scan measurement of $\n2x$}

Absolute measurement of $\n2x$ requires careful characterization of the spatiotemporal irradiance profile.
The experimental apparatus used here was identical to that used by Wahlstrand and Cruz to measure $n_2$ in fused silica \cite{wahlstrand_precise_2026},
and details, including more detailed uncertainty discussion, can be found in that paper.
The raster scan was performed using a central probe wavelength of 940~nm, for which $\Delta \omega < \Omega$ for all wavelengths within the bandwidth of the probe pulse.
The response is thus sufficiently flat in probe frequency space that spatiotemporal reconstruction of the phase shift $\Delta \phi(x,y,t)$ can be performed without distortion.

\begin{figure}[ht]
    \centering
    \includegraphics[width=14 cm]{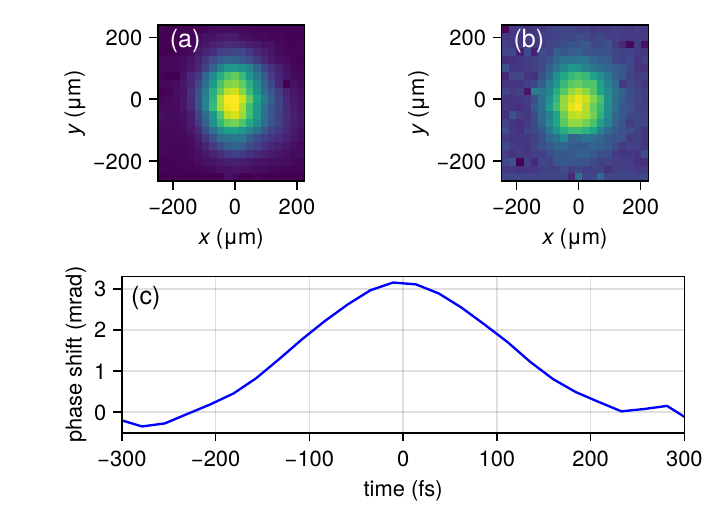}
    \caption{(a) Image of the pump spot at the sample position. (b) Phase image at the center of the pulse. (c) Time-dependent phase at the center of the spot.}
    \label{spots}
\end{figure}

An average power of $(110 \pm 7)$~$\mu$W was measured incident on the sample, with an effective repetition rate of 1.5~kHz accounting for the mechanical chopper in the pump path.
The uncertainty in average power has contributions from calibration (5~\%) and the measurement (5~$\mu$W).
The measured pump and probe profiles, the phase image, and the extracted temporal phase shift are shown in Fig.~\ref{spots}.
The spatial profile of the pump beam is well fit by a two-dimensional Gaussian with a full width at half maximum (FWHM) of $(106 \pm 4)$ $\mu$m (horizontal) by $(121 \pm 5)$ $\mu$m (vertical).
The uncertainties here have contributions from the calibration of the imaging system (3~\%) and the rest from noise in the images.
The temporal shape is also well fit by a Gaussian with FWHM ($250 \pm 8$)~fs, which we corroborated with an intensity autocorrelation \cite{wahlstrand_precise_2026}.

\section{Heat deposition by chirped pulses}

Here we derive expressions for the heat deposited by chirped pulses.
We now allow both pulses to be chirped, so that the real field inside the sample is
\begin{equation}
\mathbf{E}(t) = \hat{\mathbf{e}}_e\left[A_e(t)e^{-i\omega_e t} + \mathrm{c.c.}\right] + \hat{\mathbf{e}}_p\left[A_p(t)e^{-i\omega_p t} + \mathrm{c.c.}\right],
\label{fields}
\end{equation}
with slowly varying complex envelopes $A_j(t)$ and unit polarization vectors $\hat{\mathbf{e}}_j$.
The local irradiance of beam $j$ is
\begin{math}
I_j(t) = 2 n_0 \varepsilon_0 c\, |A_j(t)|^2 .
\end{math}
We take $\Delta\omega \equiv \omega_p - \omega_e > 0$, so that beam $p$ (905~nm in the experiment) is the higher-frequency beam and beam $e$ (1030~nm) the other.

As described in Section IIB of the main text, the optical field exerts a force
\begin{equation}
F(t) = \frac{\varepsilon_0}{2}\,\mathbf{E}(t)\cdot\frac{\partial\chi}{\partial Q}\cdot\mathbf{E}(t)
\label{force}
\end{equation}
on the phonon coordinate $Q$, which obeys
\begin{equation}
\ddot{Q} + \Gamma\dot{Q} + \Omega^2 Q = F(t).
\label{eom}
\end{equation}
Its solution is Eq.~(3) of the main text.
The Raman polarization is $\mathbf{P}_R = \varepsilon_0 (\partial\chi/\partial Q)\, Q\, \mathbf{E}$.
For the (001) surface at normal incidence, $\partial\chi/\partial Q = R_z$ with off-diagonal elements $d$.

The energy per unit volume transferred from the optical field to the medium is $\int \mathbf{E}\cdot\partial_t\mathbf{P}_R\,\dd t$.
Because $\partial\chi/\partial Q$ is symmetric,
\begin{equation*}
\int \mathbf{E}\cdot\partial_t\!\left(\varepsilon_0 \frac{\partial\chi}{\partial Q} Q\mathbf{E}\right)\dd t
= \varepsilon_0\int \left[\dot{Q}\,\mathbf{E}\cdot\frac{\partial\chi}{\partial Q}\cdot\mathbf{E} + \frac{Q}{2}\,\partial_t\!\left(\mathbf{E}\cdot\frac{\partial\chi}{\partial Q}\cdot\mathbf{E}\right)\right]\dd t
= \int F(t)\,\dot{Q}(t)\,\dd t,
\end{equation*}
where the second term was integrated by parts.
Multiplying Eq.~(\ref{eom}) by $\dot{Q}$ gives
\begin{equation*}
\frac{\dd}{\dd t}\left[\frac{1}{2}\left(\dot{Q}^2 + \Omega^2 Q^2\right)\right] + \Gamma\dot{Q}^2 = F\dot{Q},
\end{equation*}
so $(\dot{Q}^2+\Omega^2Q^2)/2$ is the phonon energy density.
The phonon amplitude starts at zero and eventually returns to zero, so all of the work is converted to heat:
\begin{equation}
U = \int_{-\infty}^{\infty} F(t)\,\dot{Q}(t)\,\dd t = \Gamma\int_{-\infty}^{\infty}\dot{Q}^2\,\dd t .
\label{U_time}
\end{equation}

\subsection{Long chirp-matched pulses}

If the envelopes vary slowly compared with $\Gamma^{-1}$ and the difference frequency $\Delta\omega$ is fixed, the phonon follows the drive adiabatically, and Eqs.~(10,11) describe the loss from beam $p$, $-\dd I_p/\dd z = \beta_R I_e I_p$.
The interaction leaves a fraction $\Delta\omega/\omega_p$ of the energy removed from beam $p$ in the lattice, so heat is deposited at the rate given by Eq.~(12) in the main text.

\subsection{General chirped pulses}
Inserting Eq.~(\ref{fields}) into Eq.~(\ref{force}) and considering that the pulse bandwidth in the experiment is insufficient for impulsive stimulated Raman scattering, we see that only difference-frequency terms near $\nu = \pm\Delta\omega$ contribute to the force. We have
\begin{math}
F(t) \simeq \varepsilon_0 \kappa\left[G(t) + G^*(t)\right],
\end{math}
where
\begin{math}
G(t) \equiv A_e^*(t)A_p(t)\,e^{-i\Delta\omega t},
\end{math}
and
\begin{math}
\kappa \equiv \hat{\mathbf{e}}_e\cdot(\partial\chi/\partial Q)\cdot\hat{\mathbf{e}}_p .
\label{F_diff}
\end{math}
With $\hat{\mathbf{e}}_e$ at angle $\theta$ from [100],
\begin{math}
\kappa^2 = d^2 s(\theta),
\end{math}
with
$s(\theta) = \sin^2(2\theta)$ for $\hat{\mathbf{e}}_p \parallel \hat{\mathbf{e}}_e$ and $s(\theta) = \cos^2(2\theta)$ for $\hat{\mathbf{e}}_p \perp \hat{\mathbf{e}}_e$.

Defining the Fourier transform of $f(t)$ as $\tilde{f}(\nu) = \int f(t)\, e^{i\nu t}\,\dd t$, a phonon lineshape function can be defined,
\begin{equation}
\chiph(\nu) \equiv \frac{1}{\Omega^2 - \nu^2 - i\Gamma\nu}.
\label{chiph}
\end{equation}
Using $\tilde{Q}(\nu) = \chiph(\nu)\tilde{F}(\nu)$, $\tilde{\dot Q}(\nu) = -i\nu\tilde{Q}(\nu)$, and Parseval's theorem,
\begin{equation}
U = \frac{1}{2\pi}\int \tilde{F}^*(\nu)\,(-i\nu)\chiph(\nu)\,\tilde{F}(\nu)\,\dd\nu
= \frac{1}{2\pi}\int_{-\infty}^{\infty} |\tilde{F}(\nu)|^2\,\nu\,\mathrm{Im}\,\chiph(\nu)\,\dd\nu .
\label{U_freq}
\end{equation}
The deposited heat is the power spectrum of the optical force weighted by the phonon absorption line shape $\nu\,\mathrm{Im}\,\chiph(\nu)$.

The spectrum $\tilde{G}(\nu)$ is concentrated near $\nu = +\Delta\omega$, and $\tilde{G^*}(\nu) = \tilde{G}^*(-\nu)$ is concentrated near $\nu = -\Delta\omega$.
Since $\Delta\omega$ is much larger than the pulse bandwidths, the two do not overlap, so $|\tilde{F}(\nu)|^2 \simeq \varepsilon_0^2 \kappa^2\left[|\tilde{G}(\nu)|^2 + |\tilde{G}(-\nu)|^2\right]$.
Since $\nu\,\mathrm{Im}\,\chiph$ is even,
\begin{equation}
U = \frac{\varepsilon_0^2\kappa^2}{\pi}\int_{-\infty}^{\infty}|\tilde{G}(\nu)|^2\,\nu\,\mathrm{Im}\,\chiph(\nu)\,\dd\nu .
\label{U_G}
\end{equation}

Defining
\begin{math}
J(t) \equiv 2n_0\varepsilon_0 c\, A_e^*(t)A_p(t)\,e^{-i\Delta\omega t},
\end{math}
where $\phi_j(t) = \arg A_j(t)$ is the temporal phase of each envelope,
the instantaneous frequency of beam $j$ is $\omega_j - \dot\phi_j$, so the instantaneous difference frequency is $\Delta\omega + \dot\phi_e - \dot\phi_p$.
Using \begin{math}
f_R\,\n2x = d^2/(4 n_0^2 c\, \Omega^2),
\label{fR}
\end{math}
$\varepsilon_0^2\kappa^2|\tilde{G}|^2 = \Omega^2 f_R \n2x\, s(\theta)\,|\tilde{J}|^2/c$, and Eq.~(\ref{U_G}) becomes the general result
\begin{equation}
U = \frac{f_R\,\n2x\,\Omega^2\, s(\theta)}{\pi c}\int_{-\infty}^{\infty}
\frac{\Gamma\nu^2}{(\Omega^2-\nu^2)^2+\Gamma^2\nu^2}\,
\left|\tilde{J}(\nu)\right|^2\dd\nu,
\end{equation}
with
\begin{equation}
\tilde{J}(\nu) = \int_{-\infty}^{\infty}\sqrt{I_e(t)I_p(t)}\;e^{i[\phi_p(t)-\phi_e(t)]}\,e^{i(\nu-\Delta\omega)t}\,\dd t .
\label{U_general}
\end{equation}
Equivalently, $\tilde{J}$ can be computed from the spectra of the two pulses.
With $\tilde{\mathcal{E}}_j(\omega) = \sqrt{2n_0\varepsilon_0 c}\int A_j(t)e^{-i\omega_j t}e^{i\omega t}\dd t$,
\begin{equation}
\tilde{J}(\nu) = \frac{1}{2\pi}\int \tilde{\mathcal{E}}_e^*(\omega)\,\tilde{\mathcal{E}}_p(\omega+\nu)\,\dd\omega ,
\label{J_spectral}
\end{equation}
which sums over all pairs of spectral components separated by $\nu$.
The heat is therefore set by how much of this spectral cross-correlation falls within the phonon line.

For $\Gamma \ll \Omega$ and $\nu$ near $+\Omega$,
\begin{equation}
\nu\,\mathrm{Im}\,\chiph(\nu) \simeq \frac{\pi}{2}\,L(\nu-\Omega),
\qquad
L(x) = \frac{1}{\pi}\frac{\Gamma/2}{x^2 + (\Gamma/2)^2},
\label{lorentz}
\end{equation}
Then
\begin{equation}
U \simeq \frac{f_R\,\n2x\,\Omega^2 s(\theta)}{2c}\int L(\nu-\Omega)\,|\tilde{J}(\nu)|^2\,\dd\nu .
\label{U_resonant}
\end{equation}
If $|\tilde{J}(\nu)|^2$ is narrow compared with $\Gamma$, the Lorentzian can be taken outside the integral.
Parseval's theorem then gives $\int|\tilde{J}|^2\dd\nu = 2\pi\int I_eI_p\,\dd t$, and Eq.~(\ref{U_general}) reduces to
\begin{equation}
U \simeq \frac{\Delta\omega}{\omega_p}\,\beta_R(\theta,\Delta\omega)\int I_e(t)I_p(t)\,\dd t ,
\label{U_cw}
\end{equation}
where $\beta_R$ is Eq.~(10) with $\lambda$ the wavelength of beam $p$.
This is the time integral of Eq.~(12).

In the experiment, we used two chirped pulses but the chirp was not exactly matched. Take
\begin{equation}
A_e(t) = A_{e0}\exp\!\left[-\left(\frac{1}{\tau_e^2}+i\xi_e\right)(t-t_d)^2\right],
\qquad
A_p(t) = A_{p0}\exp\!\left[-\left(\frac{1}{\tau_p^2}+i\xi_p\right)t^2\right],
\label{gauss_pulses}
\end{equation}
with peak irradiances $I_{j0} = 2n_0\varepsilon_0 c|A_{j0}|^2$, intensity FWHM $\sqrt{2\ln 2}\,\tau_j$, fluences $\mathcal{F}_j = \sqrt{\pi/2}\,I_{j0}\tau_j$, and delay $t_d$ of beam $e$ relative to beam $p$.
For a pulse of transform-limited width $\tau_{j0}$ stretched by group-delay dispersion $\phi_j^{(2)}$, $\tau_j^2 = \tau_{j0}^2 + 4(\phi_j^{(2)})^2/\tau_{j0}^2$ and $\xi_j = 2\phi_j^{(2)}/(\tau_{j0}^2\tau_j^2) \to 1/(2\phi_j^{(2)})$ for $|\phi_j^{(2)}| \gg \tau_{j0}^2$.

Defining
\begin{math}
a \equiv \tau_e^{-2} + \tau_p^{-2},
\end{math}
\begin{math}
\Delta \xi \equiv \xi_p - \xi_e,
\end{math}
\begin{math}
\sigma^2 \equiv a + (\Delta \xi)^2/a,
\end{math}
and
\begin{math}
\nu_c \equiv \Delta\omega + \frac{2t_d\,(\xi_e\tau_e^2 + \xi_p\tau_p^2)}{\tau_e^2 + \tau_p^2},
\end{math}
we can write
\begin{equation}
|\tilde{J}(\nu)|^2 = \frac{\pi I_{e0}I_{p0}}{\sqrt{a^2+(\Delta \xi)^2}}\,
\exp\!\left[-\frac{2t_d^2}{\tau_e^2+\tau_p^2}\right]
\exp\!\left[-\frac{(\nu-\nu_c)^2}{2\sigma^2}\right].
\label{J_gauss}
\end{equation}
The cross spectrum is a Gaussian centered at $\nu_c$ with rms width $\sigma$.
Its area, $\pi\sqrt{2\pi/a}\;I_{e0}I_{p0}\,e^{-2t_d^2/(\tau_e^2+\tau_p^2)} = 2\pi\int I_eI_p\,\dd t$, does not depend on the chirp parameters.
Chirp mismatch therefore only spreads a fixed area over a wider range of difference frequencies.

Inserting Eq.~(\ref{J_gauss}) into Eq.~(\ref{U_general}), we find
\begin{equation}
U = \frac{f_R\,\n2x\,\Omega^2 s(\theta)\, I_{e0}I_{p0}}{c\sqrt{a^2+(\Delta \xi)^2}}\,
e^{-2t_d^2/(\tau_e^2+\tau_p^2)}
\int_{-\infty}^{\infty}\frac{\Gamma\nu^2}{(\Omega^2-\nu^2)^2+\Gamma^2\nu^2}\,
e^{-(\nu-\nu_c)^2/2\sigma^2}\,\dd\nu .
\label{U_gauss_exact}
\end{equation}
For $\Gamma \ll \Omega$ and $\nu_c$ near $+\Omega$, we can derive Eq.~(12) in the main text.
\begin{equation}
U \simeq \frac{\pi\, f_R\,\n2x\,\Omega^2\, s(\theta)}{2c}\;
\frac{I_{e0}\,I_{p0}}{\sqrt{a^2+(\Delta \xi)^2}}\;
e^{-2t_d^2/(\tau_e^2+\tau_p^2)}\;
\mathrm{Re}\,\mathrm{erfcx}\!\left(-i\frac{\nu_c-\Omega + i\Gamma/2}{\sqrt{2}\,\sigma}\right).
\label{U_gauss}
\end{equation}
Equations~(\ref{U_general}) and (\ref{U_gauss}) neglect depletion of either beam.
They also ignore group-velocity walk-off.